%% file: icaif_advisor_notes.tex
\documentclass[sigconf,nonacm]{acmart}

\usepackage{tikz}
\usepackage[skip=2pt]{caption}
\AtBeginDocument{%
  \providecommand\BibTeX{{%
    \normalfont B\kern-0.5em{\scshape i\kern-0.25em b}\kern-0.8em\TeX}}}

\setcopyright{acmcopyright}
\copyrightyear{2021}
\acmYear{2021}
\acmDOI{10.1145/1122445.1122456}

\acmConference[ICAIF '21]{ICAIF '21: ACM International Conference on AI in Finance}{November 03--05, 2021}{Virtual}
\acmBooktitle{ICAIF '21: ACM  International Conference on AI in Finance, November 03--05, 2021, Virtual}
\acmPrice{15.00}
\acmISBN{978-1-4503-XXXX-X/18/06}

\begin{document}

\title[Investor Behavior Modeling by Financial Advisor Notes]{Investor Behavior Modeling by Analyzing Financial Advisor Notes: A Machine Learning Perspective}


\author{Cynthia Pagliaro}
\email{cynthia_a_paglioaro@vanguard.com}
\affiliation{%
  \institution{The Vanguard Group}
  \streetaddress{100 Vanguard Blvd}
  \city{Malvern}
  \state{PA}
  \country{USA}
  \postcode{19355}
}
\author{Dhagash Mehta}
\email{dhagash_mehta@vanguard.com}
\affiliation{%
  \institution{The Vanguard Group}
  \streetaddress{100 Vanguard Blvd}
  \city{Malvern}
  \state{PA}
  \country{USA}
  \postcode{19355}
}
\author{Han-Tai Shiao}
\email{han-tai_shiao@vanguard.com}
\affiliation{%
  \institution{The Vanguard Group}
  \streetaddress{100 Vanguard Blvd}
  \city{Malvern}
  \state{PA}
  \country{USA}
  \postcode{19355}
}
\author{Shaofei Wang}
\email{shaofei_wang2@vanguard.com}
\affiliation{%
  \institution{The Vanguard Group}
  \streetaddress{100 Vanguard Blvd}
  \city{Malvern}
  \state{PA}
  \country{USA}
  \postcode{19355}
}
\author{Luwei Xiong}
\email{luwei_xiong@vanguard.com}
\affiliation{%
  \institution{The Vanguard Group}
  \streetaddress{100 Vanguard Blvd}
  \city{Malvern}
  \state{PA}
  \country{USA}
  \postcode{19355}
}


\renewcommand{\shortauthors}{Pagliaro and Mehta, et al.}

\begin{abstract}
  Modeling investor behavior is crucial to identifying behavioral coaching opportunities for financial advisors.
  With the help of natural language processing (NLP) we analyze an unstructured (textual) dataset of financial advisors' summary notes, taken after every investor conversation, to gain first ever insights into advisor-investor interactions.
  These insights are used to predict investor needs during adverse market conditions; thus allowing advisors to coach investors and help avoid inappropriate financial decision-making.
  First, we perform topic modeling to gain insight into the emerging topics and trends.
  Based on this insight, we construct a supervised classification model to predict the probability that an advised investor will require behavioral coaching during volatile market periods.
  To the best of our knowledge, ours is the first work on exploring the advisor-investor relationship using unstructured data.
  This work may have far-reaching implications for both traditional and emerging financial advisory service models like robo-advising.
\end{abstract}





\maketitle

\section{Introduction}\label{sec:intro}
  With frequent market volatility and the need to increase the chances of meeting long and short term financial goals, financial advice services for individual as well as institutional investors have become very popular in the financial sector.
  The value of financial advice to individual as well as institutional investors is extensively studied in the literature~\cite{marsden2011value, kramer2012financial, collins2012financial, finke2013financial, vanguard2019advice}.

  The value of financial advice may not only come in the form of determining the investor's financial risk tolerance and loss aversion; selecting appropriate financial products; constructing personalized and diversified financial portfolios to suit the individual's short and long term goals; improving tax efficiency of the savings; but also comes from behavioral coaching during adversarial financial conditions (both, the investor's personal financial situations and setbacks, as well as the market fluctuations).
  Though the robo-advisors have become popular in the recent years for the cost benefits and their wide and easy accessibility, even the investor working with a robo-advisor values the possibility to reach out and interact with humans~\cite{rossi2020needs, vanguard2020value}.

  In the present work, with the help of machine learning (ML) techniques, we study advisor-investor interactions in order to predict the advised investors' needs and use these predictions to help advisors proactively coach their clients during adverse market conditions.
  In particular, we use natural language processing (NLP) techniques such as topic modeling and words embedding to extract behavioral insights from financial advisors' summary notes.

  Below, we review the existing literature on data-driven investors' modeling as well as interactions between financial advisors and their clients.

\subsection{Previous Work}
  Most ML-focused research in Finance areas focuses on market and investment performance; far less on actual investor behavior.
  In the case of the latter, there are comprehensive studies, utilizing more traditional statistical techniques that explore savings, spending and investment behavior among institutional, retail and advised investors~\cite{vanguard2020invest,vanguard2021save}.
  Though investors' behavior as well as interaction between financial advisors and investors has been investigated from traditional finance and behavioral finance~\cite{barberis2005survey, vanguard2018alpha} areas, the literature on purely data-driven investigations of these topics and, especially, using machine learning, is sparse.

  There are, however, indeed a few important research works that have laid foundations to this area: in~\cite{silva2019modeling}, a relatively large and high-frequency trading dataset of 13,000 investors of a large bank in Brazil between 2016 and 2018 was analyzed to find evidence of mean-reverting investment strategies (investors decide their investment strategies using recent past price changes), and a correlation between gender as well as academic background with mean-reverting strategy.


  In~\cite{obaid2021picture}, computer vision techniques were used to extract information from large sample of photos published in the press to create a daily investor sentiment index.
  On the other hand, a similar measure was developed based on the valence of songs that individuals listen to, which in turn captures seasonal mood swings, and was shown to be associated with a systematic pattern of mispricing correction (especially, for stocks with greater limits to arbitrage)~\cite{fernandez2020music}.

  In~\cite{rossi2020benefits}, account balance and trading dataset of 50,000 investors signed up with the personal advisory services at Vanguard was investigated using boosted decision trees and concluded that the investors that benefited the most from the advisor service were the clients with little investment experience, and those who had high cash-holdings and high trading volume pre-adoption.

  In~\cite{thompson2021know}, the authors analyzed a one year data of 23,000 investors, who were actively working with financial advisors, with the dataset including investors' demographic, recency, frequency and monetary related variables.
  The authors used unsupervised clustering technique to identify groups of investors that behaved similarly, and then showed that only demographic information of the investors, such as gender, residence region, and marital status, could not explain investors' behaviours, whereas the variables for trade and transaction frequency and volume were most informative.

  As for specifically investigating the advisor-investor interaction,~\cite{guo2015investor} studied tweets between financial advisors and investors to establish an association between investor attention and market fluctuations.
  Similarly,~\cite{zhang2015limited} using Baidu index (a search engine index corresponding to search-volume and frequency of keywords and phrases) as a proxy to investors' attention, a correlation between individual investors' attention and ChiNext stock market performance.
  In~\cite{tauni2020investor}, personality similarity between an advisor and investor, measured using an online survey questionnaire, was shown to be related to investor stock trading performance.
  In particular, it was shown that the investor-advisor similarity in terms of openness, extraversion, conscientiousness and agreeableness was positively related to the corresponding investor's stock trading performance, whereas the similarity in neuroticism negatively affected trading performance (see also, \cite{monti2014retail}).

\subsection{Our Contribution}
  In the present work, we explore the interaction between the financial advisor and investors by analyzing the unique data of advisors' summary notes prepared after every meeting with the investor.

  The notes may have free format and contain both manually written notes as well as text selected from pre-filled drop-down menus in the Customer Relationship Management (CRM) system.
  From the machine learning point of view, each of the advisor notes is an individual document with textual data.

  Machine learning techniques, particularly, NLP, have been applied with great effect in a variety of industries.
  Within financial services, NLP techniques have been used to perform sentiment analysis, topic modeling, relationship extraction, automatic summarization, etc.~for the textual data such as news articles, earning statements, manager presentations and acquisition announcements and social media posts~\cite{li2014news, schumaker2012evaluating, mishev2020evaluation, jaggi2021text, sohangir2018big}.
  However, to the best of our knowledge, the present work is the first ever work to analyze advisor-investors interaction as well as to investigate behavioral aspects from the data to identify investors in need of proactive financial and behavioral coaching.

  In this paper, we begin by applying the topic modeling technique, an NLP technique that summarizes large collection of textual information into more meaningful data elements, to the financial advisor notes data and demonstrate how this methodology can (1) help better understand the nature of advisor/client discussions and (2) convert advisor notes into a data element that can be utilized in more sophisticated behavioral models.

\section{Data Description}\label{sec:data}
  In this section, we provide details of various data sources analyzed in the present work.

\subsection{Advisor Notes}\label{sec:advisor_notes}
\begin{figure}[h]
    \centering
    \includegraphics[width=\linewidth]{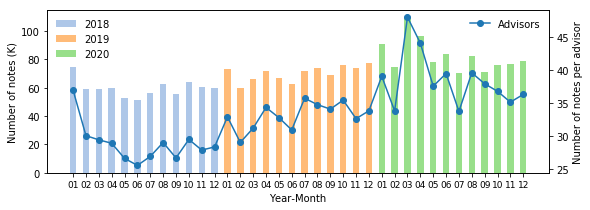}
    \caption{Total number of notes and number of notes per advisor in each month from 2018 to 2020.}
    \label{fig:monthly}
  \end{figure}

  Financial advisors conduct both regularly scheduled and ad hoc meetings with their clients.
  After each meeting, advisors document key aspects of the conversation in a CRM system within one to three days of the interaction.
  Overall, our sampled data for this paper includes about 1.5M notes from 2018 to 2020.
  Those notes were taken by more than 1,000 advisors for about 150,000 investors.
  Figure~\ref{fig:monthly} shows the monthly note total and number of notes per advisor.
  Typically, an advisor creates about 25 to 45 notes in a month.

\begin{figure}[h!]
    \centering
    \includegraphics[width=\linewidth]{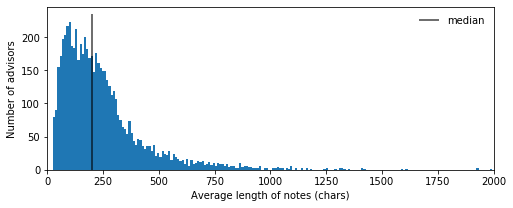}
    \caption{Histogram of the average length of notes illustrating the note-taking styles of advisors.  Nearly half of the advisors write notes shorter than 250 characters, but still a sizeable group of advisors write long and detailed notes.}
    \label{fig:advisor_style}
  \end{figure}

  While basic guidelines for documentation are outlined, advisors utilize a method or style that best works for them.
  Figure~\ref{fig:advisor_style} shows the histogram of the average length of notes created by advisors from year 2018 to 2020.
  This histogram indicates half of the advisors typically keep short and brief notes (average length is less than 250 characters), and a sizeable group of advisors keep thorough and comprehensive notes.

  Some notes are very detailed while others are simply a series of bullet points.
  We show a portion of the advisor note below for illustration,
  \begin{quote}
    \textit{Allison and Bob. Discussed MV. They don't seem too worried.  I reassured them that they are only 35\% stocks AA, and will check on regular basis.  They are now in their new assisted living facility.  They like it.}
  \end{quote}
  Some stylistic factors create a challenge for NLP.
  As above, commonly used financial phrases by advisors such as ``asset allocation'' and ``market volatility'' may be abbreviated as ``AA'' and ``MV''.
  Some advisors prefer to keep notes with bullet points and the presence of bullets and other erroneous characters should also be taken care of before most of the NLP downstream tasks, such as topic modeling.

  In general, text data preparation involves several steps: (1) using regular expression to remove non-text characters or irreverent text that can be identified through search pattern; (2) tokenize words and remove stop words; (3) creation of bigrams and trigrams; and, (4) lemmatization.
  For brevity, we skip the details about the above steps, but illustrate the data preparation and preprocessing procedures with the following examples, as illustrated in Table~\ref{tb:preprocess}.

  \begin{itemize}
    \item The first step is to tokenize the words and remove any stop words like ``and'', ``in'', or ``it'';
    \item Second, create bigrams and trigrams to capture the correct meaning, for example, the bigram ``regular-basis'' and trigram ``assisted-living-facility'' indicate the appropriate meaning of the words combined;
    \item Lemmatization is used to map variations of the word back to its simplest form.
  \end{itemize}
  Once all three steps are completed, the words can be kept as a dictionary which is ready for the modeling tasks that follow.

\subsection{Transaction Data}\label{sec:txn}
  Transaction data are a collection of trading records from investors, and the data reflect not only the trading behavior of investors but the outlook for the financial markets.
  Our transaction data include three basic elements: financial instruments (stocks, exchange traded funds, etc.), transaction types (buy, sell, etc.), and transacted amounts.
  The data is comprised from six-month trade and transaction details for about 150,000 investors.
  This data was processed, and features were derived as part of the inputs for investor behavior prediction model described in Section~\ref{sec:vit}.
  The derived features are designed to capture the investors' trading/transaction behavior, and they can be summarized into three types as shown in Table~\ref{tb:txn_features}.

  \begin{table}[t]
    \centering
    \caption{Text data preprocessing steps and examples of intermediate results in each step.}
    \label{tb:preprocess}
    \begin{tabular}{p{0.1\textwidth}p{0.325\textwidth}}
      Step & Intermediate results \\ \hline
      Tokenization, stopword removal &
        \texttt\small{{allison, bob, discussed, MV, they, dont, seem, worried, reassured, them, they, only,  stocks, AA, will, check, regular, basis, they, their, new, assisted, living, facility, they, like}} \\
      Creation of bigrams and trigrams &
        \texttt\small{{regular-basis, assisted-living-facility}} \\
      Lemmatization &
        \texttt\small{{discuss, worry, reassure, assist}}
    \end{tabular}
  \end{table}

  \begin{table}[t]
    \centering
    \caption{Summary of features derived from transaction data. Transaction features can be categorized into three types: recency, frequency, and monetary.}\label{tb:txn_features}
    \begin{tabular}{p{0.1\textwidth}p{0.325\textwidth}}
      Feature type & Description \\ \hline
      Recency
          & Capture investors' weekly trading patterns, number of trades clustered in recent days, investors' trading habits (regular or sporadic) \\
      Frequency
          & Total number in different types of trades, total number of trades in different accounts, total number of trades for the same financial instruments \\
      Monetary
          & Total amount in different types of trades, total amount of trades in different accounts, total amount of trades for the same financial instruments
    \end{tabular}
  \end{table}

\subsection{Market Data}\label{sec:vix}
  In the finance literature, the volatility of any security, fund or market trading price series is defined as the amount of variation of the price series over time and is usually measured by the standard deviation of logarithmic return.
  The volatility can be further specified as past, current or future volatility depending on the context of the volatility computation.
  There are various volatility indices available from different data vendors and institutions. We use the Chicago Board Options Exchange (CBOE) Volatility Index (VIX)~\cite{exchange2009cboe}.
  In the present work, the weekly average of VIX index was used to re-weight the significance of transactions.
  For instance, we put more emphasis on transactions that occurred during high VIX index when creating features from the transaction data.

\section{Methodology}\label{sec:method}
  In this section, we describe the methodology for both topic modeling and supervised classification used in the present work.

\subsection{Topic Modeling}

\subsubsection*{Latent Dirichlet Allocation}
  Discovering topics out of a collection of text documents is a traditional problem in machine learning that is extensively studied yet keeps posing different challenges.
  These include the variety of topics, the choice of topic modeling algorithm, the number of topics and tuning parameters used in the model.
  The topic modeling process involves both data preparation (described in Section~\ref{sec:advisor_notes}) and model generation.

  In this work, we used one of the most well-studied and popular model, called Latent Dirichlet Allocation (LDA)\cite{blei2003latent}, to identify topics from advisor notes.
  Here, with the `bag-of-words' encoding of each document, and a pre-determined number of topics, $k$, the algorithm starting with random assignment at first keeps computing and updating probability of a word being in one of the $k$ topics till it converges at a stopping criterion.

\subsubsection*{Metric for Topic Modeling}
  To find the optimal value for $k$, one scans over a range of integer values and choses when a chosen metric is optimized.
  In the present work, we have used coherent score ($C_v$)~\cite{roder2015exploring}, which is computed over sliding windows, as the metric to evaluate the quality of topic modeling.
  We used the sliding window size as default of 110 words.

\subsection{Word2Vec}
  To capture the domain specific language used by the financial advisors as well as achieving semantic similarity among words in the available corpus of text, we employed the Word2Vec embedding by starting it from scratch on our data.
  The main principle behind Word2Vec algorithm is that the meaning of a word can be inferred from its surrounding words.
  A neural-network-based algorithm is employed to learn word associations from a large text corpus and present each word as a vector.
  The word vectors, widely known as word embeddings, carry semantic information about words.
  Once trained, this model can detect synonyms for a user given word based on the training data.
  The level of semantic similarity between words can be measured by simple mathematical functions such as the cosine similarity between vectors.
  In the present work, we used the Word2Vec model that was trained on all the available notes.

  The common algorithms used for training a Word2Vec model are Continuous Bag of words (CBOW) and Skip-gram~\cite{mikolov2013distributed}.
  The Skip-gram is an unsupervised learning technique used to find the most relevant words for a given word.
  This algorithm predicts the context words for the given target word.
  Conversely, the CBOW algorithm predicts the probability of a target word while given context words.

  We first preprocessed the advisor notes with the four steps described in Section~\ref{sec:advisor_notes} and utilized Skip-gram algorithm implemented in Gensim Python library~\cite{rehurek2010software} to train the Word2Vec model and to obtain the word embeddings.
  Several parameters we chose during the Word2vec model training are summarized below.
  The dimension of the word vector is set to 100, window size to 2, and minimum word frequency to 20 to filter out the infrequent words.
  We also leveraged the negative sampling approach~\cite{goldberg2014word2vec} to speed up the training process and the number of negative samples was set to 20.

  This Word2Vec model and the corresponding word embeddings are trained from our advisor notes corpus from scratch.
  Therefore, common financial phrases and their abbreviations used by our advisors are included, and the semantic information are naturally captured in the embeddings.
  For example, the embedding vectors of \textit{market volatility} and \textit{MV} are close to each other when measured by the cosine similarity.
  The advantages of using a Word2Vec model trained from our unique corpus are (1) abbreviations can be kept for downstream tasks and that would reduce the amount of preprocessing work, (2) spelling errors, synonyms, common usage problems are automatically handled since their semantic information is more relevant, (3) the Word2Vec helps to extract targeted information (defined by keywords) from the notes as described in Section~\ref{sec:vit}.

\subsection{Classification Models}\label{sec:models}
  We employed different binary classification algorithms to train and predict the probability of an investor requiring intervention and behavioral coaching by their advisor: Logistic regression (to obtain the base line) and Gradient Boosting.

\subsubsection*{Logistic Regression}
    Logistic regression is one of the most studied and widely applied models for binary classification tasks that uses a logistic function of linear combination of input variables to model a binary target variable.
    We use this model to obtain base-line results to be matched or out-performed by more complex machine learning models.

\subsubsection*{Decision Tree}
  Decision Tree (DT) is one of the most powerful non-linear and yet interpretable machine learning algorithms that attempts to identify the decision process for the given regression or classification task from the data.
  Here, the depth (number of levels to cascade the tree to) of the tree is a hyperparameter and needs to be tuned to trade-off between bias and variance. However, DT is prone to overfitting, and sometimes its extension called Gradient Boosting Trees is preferred over DT.

\subsubsection*{Gradient Boosting}
  Gradient Boosting (GB) Trees is an ensemble learning method based on multiple ``weak'' learners, in this case, DTs: instead of constructing one DT, multiple DTs are constructed.
  Then, an aggregate of all these DTs is used to predict the final output.
  Here, the depth of each DT as well as the number of DTs both are some of the hyperparameters and need be tuned to improve learning.
  A GB evade the overfitting problem of an individual DT by ensembling multiple DTs, though the GB then is less interpretable.
  One can compute variable importance from GB that can provide interpretability up to certain extent.

\subsubsection*{Metrics for Classification}
  Since we have a binary classification task at hand, the metrics can be defined based on four basic quantities: True Positive ($f_{\text{TP}}$) which is the number of data-points from class 0 that are correctly classified by the model; True Negative ($f_{\text{TN}}$) which is the number of data-points from class 1 that are correctly classified by the model; False Negative ($f_{\text{FN}}$) which is the number of data-points from class 0 that are incorrectly classified by the model; and, False Positive ($f_{\text{FP}}$) which is the number of data-points from class 1 that are incorrectly classified by the model.

  \begin{itemize}
    \item \textbf{Accuracy}.
      Accuracy is defined as $\dfrac{ f_{\text{TP}} + f_{\text{TN}} }{ f_{\text{TP}} + f_{\text{TN}} + f_{\text{FP}} + f_{\text{FN}} }$, i.e.,~the fraction of predictions which were predicted correctly by the model.
      For imbalanced class problems though, accuracy may yield misleading conclusions.
      For such problems, the F1~score and area under the receiver operating characteristic curve (AUC-ROC) are better metrics.

    \item \textbf{F1 score}.
      The F1 score is defined as $\dfrac{ f_{\text{TP}} }{ f_{\text{TP}} + (f_{\text{FP}} + f_{\text{FN}}) / 2}$, which takes values between 0 and 1, where F1 equals 1 meaning a perfect classification.
      To take the highly imbalanced data, we used the weighted F1 score which weighs the contributions of each class proportional to the respective number of data-points.

    \item \textbf{AUC-ROC}.
      The receiver operating characteristic (ROC) curve is a plot of the fraction of the true positive rate (TPR) \textit{vs}~the fraction of the false positive rate (FPR) and it yields the performance of the binary classifier as a function of the discrimination probability threshold.
      The area under the ROC curve (AUC-ROC) is interpreted the probability that a classifier will rank a randomly chosen data-point of class 0 higher than a randomly chosen data-point of class 1.
  \end{itemize}

\subsubsection*{Cross-Validation and Hyperparameter Optimization}
  To identify the best hyperparameter point as well as to prevent the model from overfitting the data, we used cross validation for the RF and GB models.
  Here, the $k$-fold cross validation method was used where the training set is spilt into $k$ smaller sets.
  The given model is then trained on $k-1$ of the folds and validated on the remaining fold.
  We used stratified 5-fold cross validation which ensured that all the folds the same percentage of samples of each target class.

\section{Results}\label{sec:results}
  The objective of this study was two-fold: First, determine the main themes of the advisor-investor interaction. Second, determine if/how these themes change over market conditions.
  In this section, we begin by peeking into the advisor notes and providing basic statistics of the data.
  Then, we obtain a high-level comprehension about the given textual data through topic modeling. With the insights derived from topic modeling, the second use case is to extract targeted information, augment with transaction data, and build a model for investor behavior prediction.

\subsection{First Insights into Advisor Notes}\label{sec:topic}
  We chose the notes from March~2019 and March~2020 because they represent two different financial market conditions, the former being fairly ``typical'' whereas during the latter particularly volatile due to the Covid-19 pandemic.
  The total number of notes in those two months was about 150,000.

\begin{figure}[t]
    \centering
    \includegraphics[width=\linewidth]{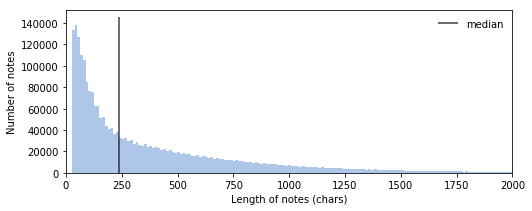}
    \caption{Histogram of the note lengths. More than 50\% of the notes is shorter than 250 characters; however, a note can be longer than 1500 characters as well.}
    \label{fig:note_length}
  \end{figure}

  The histogram of the length of notes shown in Figure~\ref{fig:note_length} highlights this variation in note taking style.
  On average, a note contains less than 250 characters excluding all whitespaces.
  Hence, the length of a note is about the same as a long tweet (according to Twitter's 280-character limit).
  While some advisors take detailed notes, most of them are keeping the notes short and brief.

  \subsubsection*{Topic Modeling}
\begin{figure}[h]
    \centering
    \includegraphics[width=\linewidth]{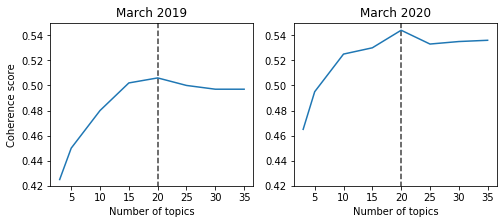}
    \caption{The coherence scores as a function of numbers of topics. The optimal number of topics for the LDA models is 20 for both periods.}
    \label{fig:coherence}
  \end{figure}

  To determine the optimal number of topics during the LDA modeling stage, we used the coherence score ($C_v$) to choose the number of topics, i.e.~, the number of topics corresponding to the highest value of $C_v$ was considered as the optimal number of topics.
  The coherence scores for a range of number of topics are shown in Figure~\ref{fig:coherence}.
  For the notes from March 2019 and March 2020, clearly the optimal number of topics should be 20.

  In Table~\ref{tb:lda_results}, we list the 20 topics for 2019 notes.
  We also show the percentage of notes assigned to (or related to) each topic and the topic keywords associated with the topic.
  For instance, Topic \textit{ Quarterly review for asset allocation and rebalance} represents around 13.3\% of notes defined by the keywords: ``send'', ``target'', ``allocation'', ``current'', ``email'', ``rebalance'', ``complete'', ``quarterly'', etc.

  In addition to the raw topics from the LDA model, we also include the topic labels in Table~\ref{tb:lda_results}.
  These topic labels were inferred from the keywords based on our best judgement, and understandably the topic label creation is a subjective process.

\begin{figure}[h]
    \centering
    \input{themes.tex}
    \caption{Topic themes comparison between March 2019 and March 2020 data. The percentages indicate the proportion of notes for the topic themes. The \textit{Inheritance} theme is not shown for March 2020 due to low number of notes.  A distinct shift can be observed in two themes: \textit{Financial planning} and \textit{Market discussion}.}
    \label{fig:themes}
  \end{figure}
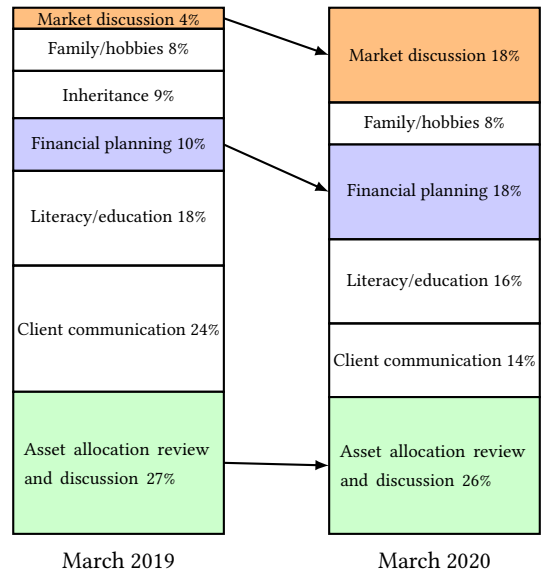

  A quick scan of the topic labels shows some commonality in the individual topics, particularly to a subject matter expert.
  In this case, a consensus view among independent subject matter experts was conducted to further consolidate the topics based on the topic keywords and given labels, thus reducing the number of topics from 20 to 7.
  This process was performed for the modeling results for both periods and the consolidation is shown in Figure~\ref{fig:themes}.

  During both periods, \textit{Asset allocation review and discussions} is the most common theme, representing 27\% and 26\% of the advisor notes, respectively.
  Given that the primary purpose of regularly scheduled meetings or advisor-investor interactions is to review investors' financial portfolio, \textit{Asset allocation review and discussions} is expected to be the major theme in both periods and those periods excluded from the topic modeling.

  \begin{table*}[ht!]
    \centering
    \caption{The 20 topics from LDA for March 2019 data including the percentage of notes, keywords, and the assigned label.}\label{tb:lda_results}
    \begin{tabular}{p{0.04\textwidth}p{0.63\textwidth}p{0.27\textwidth}}
      Notes & Keywords & Topic labels \\ \hline
      13.3\%
        & send, target, allocation, current, email, rebalance, complete, quarterly, imh, trade
        & Quarterly review for asset allocation and rebalance \\
      7.4\%
        &	schedule, appointment, call, follow, question, email, appt, message, leave, advisor
        & Client communications and schedule appointments \\
      6.5\%
        & plan, discuss, service, update, step, consent, strategy, implement, discussion, accept
        & Discuss advisory methodology, possibly client onboarding \\
      6.0\%
        & account, manage, transfer, move, taxable, open, joint, fee, trade, process
        & Upgrade inherited account \\
      5.9\%
        & fund, cash, spend, invest, money, spending, balance, emergency, position, agree
        & Spending/saving/investing strategies \\
      5.8\%
        & stock, bond, fund, gain, total, recommend, portfolio, index, hold, sell
        & Investment/portfolio discussions \\
      5.7\%
        & tax, make, contribution, taxis, distribution, state, amount, taxable, conversion, speak
        & Distributions and taxation \\
      5.6\%
        & send, set, confirm, check, bank, form, acct, receive, add, request
        & Client administration \\
      4.8\%
        & asset, plan, allocation, flow, tolerance, risk, cash, relationship, step, build
        & Intro call-asset allocation \\
      4.4\%
        & year, work, live, daughter, son, travel, kid, trip, enjoy, family
        & Family/hobbies \\
      4.4\%
        & client, update, account, conversation, discussion, profile, set, result, video, service
        & Client administration \\
      3.9\%
        & call, speak, today, week, back, time, give, day, number, end
        & Client administration \\
      3.7\%
        & investment, manage, portfolio, advisor, advice, strategy, management, financial, interested
        & Retirement planning \\
      3.6\%
        & trust, planning, cost, estate, care, insurance, mention, life, pass, health
        & Financial plan \\
      3.6\%
        & market, portfolio, return, performance, explain, volatility, concern, risk, understand, time
        & Market discussion \\
      3.4\%
        & home, pay, sell, move, money, buy, purchase, put, year, mortgage
        & Personal/family/life events \\
      3.3\%
        & year, expense, income, month, cover, increase, pay, amount, cost, start
        & Inheritance \\
      3.1\%
        & retirement, plan, work, age, retire, goal, year, saving, time, benefit
        &	Retirement planning \\
      3.1\%
        & review, discuss, change, show, spending, tool, future, dynamic, outlook, add
        & Account review \\
      2.8\%
        & talk, make, good, time, move, thing, feel, share, point, give
        & Set up new account \\
    \end{tabular}
  \end{table*}

  However, there is a distinct shift in two themes as highlighted in Figure~\ref{fig:themes}.
  The proportion of advisor notes containing the \textit{Financial planning} theme increases from 10\% to 18\%, and the \textit{Market discussion} from 4\% to 18\%.
  These significant shifts in topic themes clearly were driven by the financial market conditions.
  Further, the percentages for \textit{Family/hobbies} are about the same for both periods potentially indicating the standard conversation starters as well as conversations related to personal situations and goals.

  The findings through topic modeling and topic theme identification help us better understand the nature of the financial advisory interaction.
  Moreover, using advisor notes to study the qualitative value that an advisory service can provide also helps us evaluate the efficacy of behavioral coaching.

\subsection{Investor Behavior Prediction}\label{sec:vit}
  While performing topic modeling using advisor notes gave unique insights into and understanding about the financial advisory interaction, we further focused on specific topic(s) such as market discussion, financial planning, etc.~and utilizing them to develop a model which predicts the behavior of the investor and the timing for the advisor to proactively intervene and provide behavioral coaching.
  In other words, we further used NLP to extract targeted information from advisor notes by building a model that predicts investor behavior, i.e.,~if the investor would move all asset into cash during a volatile financial market period.

\subsubsection*{Information Extraction using Word2Vec}
  Since the model we are after is supposed to predict investor behavior in a volatile market, one natural consideration for the model inputs would be around (1) if an investor has revealed their concern for a volatile market; (2) if an investor is particularly anxious about market downturn; and (3) if an investor is sensitive and/or fear about market fluctuation.

  Such information can be identified and tagged within advisor notes using the aforementioned NLP technique.
  For example, the word ``volatility'' representing the topic about market volatility can be used to determine if market volatility was a topic in advisor notes.
  The process of identifying and tagging a given word in texts data can be more than just search verbatim.

  One common approach is to convert all the tokenized words into their Word2Vec embedding, a vector representation carrying semantic relationship about words.
  Then the closeness between words in advisor notes and the word representing a topic can be measured by the cosine similarity between the corresponding vectors in the embedding.
  That is, when the similarity measure for a word and ``volatility'' is above a particular threshold, this word is considered close to ``volatility'' in the semantic sense, and we can apply a tag to this word, representing the topic has been located.

  More concretely, we show the top ten words similar to ``volatility'' from our Word2Vec model in Table~\ref{tb:similar_words}.
  Some of them are phrases that can be used in place of ``volatility'', and some of them are misspelled (indicated by asterisk).
  If we chose 0.84 as the minimal level of similarity measure, those ten words can be identified and tagged if occurred in the advisor notes.
  The proper minimal level of similarity measure (or threshold) can vary from one topic to the other.
  However, we empirically set the threshold as 0.7 for all topics during the information extraction phase.

  \begin{table}[t]
    \centering
    \caption{Ten similar words to \textit{volatility} and their cosine similarities from our Word2Vec model.}\label{tb:similar_words}
    \begin{tabular}{lc}
      Word & Cosine similarity \\ \hline
      *volatilty          & 0.92586 \\
      market-volatility   & 0.89295 \\
      up-down             & 0.87184 \\
      downturn            & 0.86821 \\
      *volatiltiy         & 0.85863 \\
      *volitility         & 0.85847 \\
      recent-volatility   & 0.85490 \\
      increased-volatility & 0.84646 \\
      *volatiliy          & 0.84449 \\
      turbulence          & 0.84355
    \end{tabular}
  \end{table}

\subsubsection*{Feature Engineering}
  Once the words about a topic are identified and tagged, we performed feature engineering on those tags.
  Here we show a few feature engineering considerations for our prediction task. The features from advisor notes can be (1) how often does a tag appear in the notes, (2) is the occurrence of tag significantly higher when normalized by the number of notes for a given client, (3) does the tag occur more often recently, (4) are the underlying similarity measures of tags closer to one (hinting that words are fully matched with the topic)?

  The topics considered for this case study fall into two categories: one is about market volatility (represented by keywords ``market'' and ``volatility''), and the other is about investors' peace of mind (represented by keywords ``sensitive'', ``concern'', ``panic'', etc.).

  The features created from advisor notes constitute one part of the inputs for our model predicting investor behavior.
  The rest of the inputs are the features derived from the transaction data.
  Those features include number of days making trades, weekly trade frequency weighted with VIX price (described in Section~\ref{sec:vix}), the frequencies and amounts in different types of trades, week-to-week transaction pattern, etc.

\subsubsection*{Model Performance}
  The dataset we prepared for this case study includes features and labels for about 150,000 investors.
  For each investor, there are more than 100 features characterizing market volatility and peace of mind topics from advisor notes and about 50 features capturing trading behavior from transaction data.

  Further, the label is a binary value indicating if the investor chose to move all assets into cash.
  Since only a small fraction of investors in our dataset move all assets to cash, the dataset is highly imbalanced, posing a challenge for predictive modeling.
  To mitigate this issue, we have used various metrics such as weighted F1 score and AUC-ROC as discussed in Section \ref{sec:models}.

\begin{figure}[t]
    \centering
    \includegraphics[width=\linewidth]{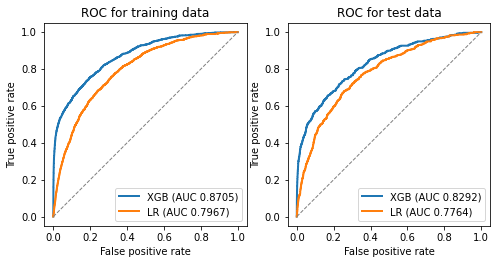}
    \caption{The ROC curves for Logistic regression and Gradient Boosting models from a representative fold. The corresponding AUCs are indicated in the legend.}
    \label{fig:roc}
  \end{figure}

  Below, we discuss the results and performance of the models described in Section~\ref{sec:models}.
  The averages of accuracy and weighted F1 scores for three classification models are shown in Table~\ref{tb:f1}.
  Noticeably, the GB model outperforms the Logistic regression model in both metrics.
  In practice, the AUC-ROC is an alternative for imbalanced classification problem.
  We show the ROC curves for training and test data in Figure~\ref{fig:roc}.
  Since we employed stratified 5-fold cross validation, Figure~\ref{fig:roc} is the result from just one representative fold.

  \begin{table}[t]
    \centering
    \caption{The average accuracies and weighted F1 scores for Logistic regression and Gradient Boosting models.}\label{tb:f1}
    \begin{tabular}{l|ll}
      Metrics & Logistic & GB \\ \hline
      Acc (train) & 0.6913 & 0.9871 \\
      Acc (test)  & 0.6908 & 0.9870 \\
      F1 (train)  & 0.8052 & 0.9813 \\
      F1 (test)   & 0.8049 & 0.9808
    \end{tabular}
  \end{table}

\subsubsection*{Feature Importance}
\begin{figure}[h]
    \centering
    \includegraphics[width=\linewidth]{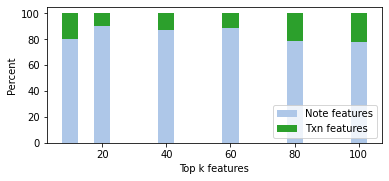}
    \caption{The percentages of features from advisor notes and transaction data for the top $k$ features.}
    \label{fig:feat}
  \end{figure}

  We also examine the feature importance based on the modeling results.
  The features are ranked according to the absolute values of the coefficients from Logistic regression model.
  In Figure~\ref{fig:feat}, we show the percent of features derived from advisor notes when consider the top $k$ features.
  Among the top 10 features, eight are from advisor notes and two from transaction data.
  Clearly, features from advisor notes are more important in the model.
  To validate which data source is more relevant for investor behavior prediction, we also trained models using advisor note features or transaction features alone.
  The results show that advisor notes give better prediction performance than the transaction data.
  However, combining both can provide a boost in the model performance.
  Details are omitted here to conserve space.

\section{Conclusions}\label{sec:conclusions}
  In addition to improved investment outcomes and the attainment of financial goals, the value of advice includes a sense of financial well-being achieved through behavioral coaching.
  Such guidance is particularly valuable during periods of market volatility.
  Of general interest to the financial advisory industry is a deeper understanding of the nature and impact of the advisor-investor relationship.
  A detailed understanding helps both advisors and service providers improve investor outcomes.

  In the present work, we used the advisor notes to explore the advisor-investor interaction and build a model to predict investor behavior.
  We used LDA to identify important topics emerging from the unstructured data.
  The final list of topics included: ``Asset allocation review and discussion'', ``Client communications'', ``Financial literacy education'', ``Financial planning'', ``Inheritance'', ``Family/hobbies'', and ``Market discussions''.
  Two topics: ``Financial planning'' and ``Market discussions'' became prominent during March 2020 when the Covid-19 related market volatility was at its peak.

  We then trained a Word2Vec model using the advisor notes data to learn an embedded representation of the data that captures domain specific semantic similarity.
  Using the insights learned from the topic modeling analysis, investor transaction data and a market volatility index (VIX), we constructed a supervised classification model to predict the investors who may require behavioral coaching from a financial advisor.

  While an important source of information, advisor notes do have limitations.
  Advisors may tend to summarize conversations in neutral or positive tones, so this type of analysis may not lend itself to sentiment analysis.
  The notes may also be represent a selective part of the conversation thus not capturing its true intent.
  However, the methodology proposed in the present work may be applied to a wide range of data sources also captured during the advisor-investor interaction.
  These include; survey responses, comments fields, emails, texts, phone call transcripts and social media posts.

  In the future, it would be of interest to apply other machine learning techniques to further understand the impact of financial advice on investors' financial decisions among all advisory models.

\section{Acknowledgments}
  The work presented here is a result of a pure and exploratory research work by the authors, and the authors are solely responsible for any mistakes and not The Vanguard Group.

  Notes: All investing is subject to risk, including the possible loss of the money you invest.
  Diversification does not ensure a profit or protect against a loss.

  \copyright2021 The Vanguard Group, Inc. All rights reserved.


\bibliographystyle{ACM-Reference-Format}
\bibliography{advisor_notes}



\end{document}

%% file: themes.tex
\begin{tikzpicture}[thick,>=latex,scale=0.7]
    \tikzstyle{block} =[draw,rectangle,minimum width=2em,minimum height=2em]

    \draw [fill=orange!50] (0,0) rectangle node[midway] {\footnotesize{Market discussion 4\%}} (4,-0.4);
    \draw [] (0,-0.4) rectangle node[midway] {\footnotesize{Family/hobbies 8\%}} (4,-1.2);
    \draw [] (0,-1.2) rectangle node[midway] {\footnotesize{Inheritance 9\%}} (4,-2.1);
    \draw [fill=blue!20] (0,-2.1) rectangle node[midway] {\footnotesize{Financial planning 10\%}} (4,-3.1);
    \draw [] (0,-3.1) rectangle node[midway] {\footnotesize{Literacy/education 18\%}} (4,-4.9);
    \draw [] (0,-4.9) rectangle node[midway] {\footnotesize{Client communication 24\%}} (4,-7.3);
    \draw [fill=green!20] (0,-7.3) rectangle node[midway, text width=2.5cm] {\footnotesize{Asset allocation review and discussion 27\%}} (4,-10);

    \draw [fill=orange!50] (6,0) rectangle node[midway] {\footnotesize{Market discussion 18\%}} (10,-1.8);
    \draw [] (6,-1.8) rectangle node[midway] {\footnotesize{Family/hobbies 8\%}} (10,-2.6);
    \draw [fill=blue!20] (6,-2.6) rectangle node[midway] {\footnotesize{Financial planning 18\%}} (10,-4.4);
    \draw [] (6,-4.4) rectangle node[midway] {\footnotesize{Literacy/education 16\%}} (10,-6.0);
    \draw [] (6,-6.0) rectangle node[midway] {\footnotesize{Client communication 14\%}} (10,-7.4);
    \draw [fill=green!20] (6,-7.4) rectangle node[midway, text width=2.5cm] {\footnotesize{Asset allocation review and discussion 26\%}} (10,-10);

 	\draw[->] (4,-0.2) -- (6,-0.9);
 	\draw[->] (4,-2.6) -- (6,-3.5);
 	\draw[->] (4,-8.65) -- (6,-8.7);

 	\node[] at (2,-10.5) {March 2019};
 	\node[] at (8,-10.5) {March 2020};
\end{tikzpicture}

%% file: icaif_advisor_notes.bbl

\begin{thebibliography}{31}


\ifx \showCODEN    \undefined \def \showCODEN     #1{\unskip}     \fi
\ifx \showDOI      \undefined \def \showDOI       #1{#1}\fi
\ifx \showISBNx    \undefined \def \showISBNx     #1{\unskip}     \fi
\ifx \showISBNxiii \undefined \def \showISBNxiii  #1{\unskip}     \fi
\ifx \showISSN     \undefined \def \showISSN      #1{\unskip}     \fi
\ifx \showLCCN     \undefined \def \showLCCN      #1{\unskip}     \fi
\ifx \shownote     \undefined \def \shownote      #1{#1}          \fi
\ifx \showarticletitle \undefined \def \showarticletitle #1{#1}   \fi
\ifx \showURL      \undefined \def \showURL       {\relax}        \fi
\providecommand\bibfield[2]{#2}
\providecommand\bibinfo[2]{#2}
\providecommand\natexlab[1]{#1}
\providecommand\showeprint[2][]{arXiv:#2}

\bibitem[\protect\citeauthoryear{Barberis and Thaler}{Barberis and
  Thaler}{2005}]%
        {barberis2005survey}
\bibfield{author}{\bibinfo{person}{Nicholas Barberis} {and}
  \bibinfo{person}{Richard Thaler}.} \bibinfo{year}{2005}\natexlab{}.
\newblock \bibinfo{booktitle}{\emph{A survey of behavioral finance}}.
\newblock \bibinfo{publisher}{Princeton University Press}.
\newblock


\bibitem[\protect\citeauthoryear{Bennyhoff, Kinniry, and DiJoseph}{Bennyhoff
  et~al\mbox{.}}{2018}]%
        {vanguard2018alpha}
\bibfield{author}{\bibinfo{person}{Donald~G. Bennyhoff},
  \bibinfo{person}{Francis~M. Kinniry}, {and} \bibinfo{person}{Michael~A.
  DiJoseph}.} \bibinfo{year}{2018}\natexlab{}.
\newblock \showarticletitle{The evolution of Vanguard Advisor's Alpha}.
\newblock \bibinfo{journal}{\emph{Vanguard research}} (\bibinfo{year}{2018}).
\newblock
\newblock
\shownote{Available at advisors.vanguard.com.}


\bibitem[\protect\citeauthoryear{Blei, Ng, and Jordan}{Blei
  et~al\mbox{.}}{2003}]%
        {blei2003latent}
\bibfield{author}{\bibinfo{person}{David~M Blei}, \bibinfo{person}{Andrew~Y
  Ng}, {and} \bibinfo{person}{Michael~I Jordan}.}
  \bibinfo{year}{2003}\natexlab{}.
\newblock \showarticletitle{Latent dirichlet allocation}.
\newblock \bibinfo{journal}{\emph{the Journal of machine Learning research}}
  \bibinfo{volume}{3} (\bibinfo{year}{2003}), \bibinfo{pages}{993--1022}.
\newblock


\bibitem[\protect\citeauthoryear{Collins}{Collins}{2012}]%
        {collins2012financial}
\bibfield{author}{\bibinfo{person}{J~Michael Collins}.}
  \bibinfo{year}{2012}\natexlab{}.
\newblock \showarticletitle{Financial advice: A substitute for financial
  literacy?}
\newblock \bibinfo{journal}{\emph{Financial Services Review}}
  \bibinfo{volume}{21}, \bibinfo{number}{4} (\bibinfo{year}{2012}),
  \bibinfo{pages}{307}.
\newblock


\bibitem[\protect\citeauthoryear{Exchange}{Exchange}{2009}]%
        {exchange2009cboe}
\bibfield{author}{\bibinfo{person}{Chicago Board~Options Exchange}.}
  \bibinfo{year}{2009}\natexlab{}.
\newblock \showarticletitle{The CBOE volatility index-VIX}.
\newblock \bibinfo{journal}{\emph{White Paper}} (\bibinfo{year}{2009}),
  \bibinfo{pages}{1--23}.
\newblock


\bibitem[\protect\citeauthoryear{Fernandez-Perez, Garel, and
  Indriawan}{Fernandez-Perez et~al\mbox{.}}{2020}]%
        {fernandez2020music}
\bibfield{author}{\bibinfo{person}{Adrian Fernandez-Perez},
  \bibinfo{person}{Alexandre Garel}, {and} \bibinfo{person}{Ivan Indriawan}.}
  \bibinfo{year}{2020}\natexlab{}.
\newblock \showarticletitle{Music sentiment and stock returns}.
\newblock \bibinfo{journal}{\emph{Economics Letters}}  \bibinfo{volume}{192}
  (\bibinfo{year}{2020}), \bibinfo{pages}{109260}.
\newblock


\bibitem[\protect\citeauthoryear{Finke}{Finke}{2013}]%
        {finke2013financial}
\bibfield{author}{\bibinfo{person}{Michael Finke}.}
  \bibinfo{year}{2013}\natexlab{}.
\newblock \showarticletitle{Financial Advice: Does it make a difference?}
\newblock \bibinfo{journal}{\emph{The market for retirement financial advice}}
  (\bibinfo{year}{2013}), \bibinfo{pages}{229--48}.
\newblock


\bibitem[\protect\citeauthoryear{Goldberg and Levy}{Goldberg and Levy}{2014}]%
        {goldberg2014word2vec}
\bibfield{author}{\bibinfo{person}{Yoav Goldberg} {and} \bibinfo{person}{Omer
  Levy}.} \bibinfo{year}{2014}\natexlab{}.
\newblock \showarticletitle{word2vec Explained: deriving Mikolov et al.'s
  negative-sampling word-embedding method}.
\newblock \bibinfo{journal}{\emph{arXiv preprint arXiv:1402.3722}}
  (\bibinfo{year}{2014}).
\newblock


\bibitem[\protect\citeauthoryear{Guo, Finke, and Mulholland}{Guo
  et~al\mbox{.}}{2015}]%
        {guo2015investor}
\bibfield{author}{\bibinfo{person}{Tao Guo}, \bibinfo{person}{Michael Finke},
  {and} \bibinfo{person}{Barry Mulholland}.} \bibinfo{year}{2015}\natexlab{}.
\newblock \showarticletitle{Investor attention and advisor social media
  interaction}.
\newblock \bibinfo{journal}{\emph{Applied Economics Letters}}
  \bibinfo{volume}{22}, \bibinfo{number}{4} (\bibinfo{year}{2015}),
  \bibinfo{pages}{261--265}.
\newblock


\bibitem[\protect\citeauthoryear{Jaggi, Mandal, Narang, Naseem, and
  Khushi}{Jaggi et~al\mbox{.}}{2021}]%
        {jaggi2021text}
\bibfield{author}{\bibinfo{person}{Mukul Jaggi}, \bibinfo{person}{Priyanka
  Mandal}, \bibinfo{person}{Shreya Narang}, \bibinfo{person}{Usman Naseem},
  {and} \bibinfo{person}{Matloob Khushi}.} \bibinfo{year}{2021}\natexlab{}.
\newblock \showarticletitle{Text Mining of Stocktwits Data for Predicting Stock
  Prices}.
\newblock \bibinfo{journal}{\emph{Applied System Innovation}}
  \bibinfo{volume}{4}, \bibinfo{number}{1} (\bibinfo{year}{2021}),
  \bibinfo{pages}{13}.
\newblock


\bibitem[\protect\citeauthoryear{Kramer}{Kramer}{2012}]%
        {kramer2012financial}
\bibfield{author}{\bibinfo{person}{Marc~M Kramer}.}
  \bibinfo{year}{2012}\natexlab{}.
\newblock \showarticletitle{Financial advice and individual investor portfolio
  performance}.
\newblock \bibinfo{journal}{\emph{Financial Management}} \bibinfo{volume}{41},
  \bibinfo{number}{2} (\bibinfo{year}{2012}), \bibinfo{pages}{395--428}.
\newblock


\bibitem[\protect\citeauthoryear{Li, Xie, Chen, Wang, and Deng}{Li
  et~al\mbox{.}}{2014}]%
        {li2014news}
\bibfield{author}{\bibinfo{person}{Xiaodong Li}, \bibinfo{person}{Haoran Xie},
  \bibinfo{person}{Li Chen}, \bibinfo{person}{Jianping Wang}, {and}
  \bibinfo{person}{Xiaotie Deng}.} \bibinfo{year}{2014}\natexlab{}.
\newblock \showarticletitle{News impact on stock price return via sentiment
  analysis}.
\newblock \bibinfo{journal}{\emph{Knowledge-Based Systems}}
  \bibinfo{volume}{69} (\bibinfo{year}{2014}), \bibinfo{pages}{14--23}.
\newblock


\bibitem[\protect\citeauthoryear{Madamba, Pagliaro, and Utkus}{Madamba
  et~al\mbox{.}}{2020}]%
        {vanguard2020value}
\bibfield{author}{\bibinfo{person}{Anna Madamba}, \bibinfo{person}{Cynthia
  Pagliaro}, {and} \bibinfo{person}{Stephen~P. Utkus}.}
  \bibinfo{year}{2020}\natexlab{}.
\newblock \showarticletitle{The value of advice: Assessing the role of
  emotions}.
\newblock \bibinfo{journal}{\emph{Vanguard research}} (\bibinfo{year}{2020}).
\newblock
\newblock
\shownote{Available at institutional.vanguard.com.}


\bibitem[\protect\citeauthoryear{Marsden, Zick, and Mayer}{Marsden
  et~al\mbox{.}}{2011}]%
        {marsden2011value}
\bibfield{author}{\bibinfo{person}{Mitchell Marsden},
  \bibinfo{person}{Cathleen~D Zick}, {and} \bibinfo{person}{Robert~N Mayer}.}
  \bibinfo{year}{2011}\natexlab{}.
\newblock \showarticletitle{The value of seeking financial advice}.
\newblock \bibinfo{journal}{\emph{Journal of family and economic issues}}
  \bibinfo{volume}{32}, \bibinfo{number}{4} (\bibinfo{year}{2011}),
  \bibinfo{pages}{625--643}.
\newblock


\bibitem[\protect\citeauthoryear{Mikolov, Sutskever, Chen, Corrado, and
  Dean}{Mikolov et~al\mbox{.}}{2013}]%
        {mikolov2013distributed}
\bibfield{author}{\bibinfo{person}{Tomas Mikolov}, \bibinfo{person}{Ilya
  Sutskever}, \bibinfo{person}{Kai Chen}, \bibinfo{person}{Greg Corrado}, {and}
  \bibinfo{person}{Jeffrey Dean}.} \bibinfo{year}{2013}\natexlab{}.
\newblock \showarticletitle{Distributed representations of words and phrases
  and their compositionality}.
\newblock \bibinfo{journal}{\emph{arXiv preprint arXiv:1310.4546}}
  (\bibinfo{year}{2013}).
\newblock


\bibitem[\protect\citeauthoryear{Mishev, Gjorgjevikj, Vodenska, Chitkushev, and
  Trajanov}{Mishev et~al\mbox{.}}{2020}]%
        {mishev2020evaluation}
\bibfield{author}{\bibinfo{person}{Kostadin Mishev}, \bibinfo{person}{Ana
  Gjorgjevikj}, \bibinfo{person}{Irena Vodenska}, \bibinfo{person}{Lubomir~T
  Chitkushev}, {and} \bibinfo{person}{Dimitar Trajanov}.}
  \bibinfo{year}{2020}\natexlab{}.
\newblock \showarticletitle{Evaluation of sentiment analysis in finance: from
  lexicons to transformers}.
\newblock \bibinfo{journal}{\emph{IEEE Access}}  \bibinfo{volume}{8}
  (\bibinfo{year}{2020}), \bibinfo{pages}{131662--131682}.
\newblock


\bibitem[\protect\citeauthoryear{Monti, Pelligra, Martignon, and Berg}{Monti
  et~al\mbox{.}}{2014}]%
        {monti2014retail}
\bibfield{author}{\bibinfo{person}{Marco Monti}, \bibinfo{person}{Vittorio
  Pelligra}, \bibinfo{person}{Laura Martignon}, {and} \bibinfo{person}{Nathan
  Berg}.} \bibinfo{year}{2014}\natexlab{}.
\newblock \showarticletitle{Retail investors and financial advisors: New
  evidence on trust and advice taking heuristics}.
\newblock \bibinfo{journal}{\emph{Journal of Business Research}}
  \bibinfo{volume}{67}, \bibinfo{number}{8} (\bibinfo{year}{2014}),
  \bibinfo{pages}{1749--1757}.
\newblock


\bibitem[\protect\citeauthoryear{Obaid and Pukthuanthong}{Obaid and
  Pukthuanthong}{2021}]%
        {obaid2021picture}
\bibfield{author}{\bibinfo{person}{Khaled Obaid} {and} \bibinfo{person}{Kuntara
  Pukthuanthong}.} \bibinfo{year}{2021}\natexlab{}.
\newblock \showarticletitle{A picture is worth a thousand words: Measuring
  investor sentiment by combining machine learning and photos from news}.
\newblock \bibinfo{journal}{\emph{Journal of Financial Economics}}
  (\bibinfo{year}{2021}).
\newblock


\bibitem[\protect\citeauthoryear{Pagliaro and Utkus}{Pagliaro and
  Utkus}{2019}]%
        {vanguard2019advice}
\bibfield{author}{\bibinfo{person}{Cynthia Pagliaro} {and}
  \bibinfo{person}{Stephen~P. Utkus}.} \bibinfo{year}{2019}\natexlab{}.
\newblock \showarticletitle{Assessing the Value of Advice}.
\newblock \bibinfo{journal}{\emph{Vanguard research}} (\bibinfo{year}{2019}).
\newblock
\newblock
\shownote{Available at institutional.vanguard.com.}


\bibitem[\protect\citeauthoryear{Rehurek and Sojka}{Rehurek and Sojka}{2010}]%
        {rehurek2010software}
\bibfield{author}{\bibinfo{person}{Radim Rehurek} {and} \bibinfo{person}{Petr
  Sojka}.} \bibinfo{year}{2010}\natexlab{}.
\newblock \showarticletitle{Software framework for topic modelling with large
  corpora}. In \bibinfo{booktitle}{\emph{In Proceedings of the LREC 2010
  workshop on new challenges for NLP frameworks}}. Citeseer.
\newblock


\bibitem[\protect\citeauthoryear{R{\"o}der, Both, and Hinneburg}{R{\"o}der
  et~al\mbox{.}}{2015}]%
        {roder2015exploring}
\bibfield{author}{\bibinfo{person}{Michael R{\"o}der}, \bibinfo{person}{Andreas
  Both}, {and} \bibinfo{person}{Alexander Hinneburg}.}
  \bibinfo{year}{2015}\natexlab{}.
\newblock \showarticletitle{Exploring the space of topic coherence measures}.
  In \bibinfo{booktitle}{\emph{Proceedings of the eighth ACM international
  conference on Web search and data mining}}. \bibinfo{pages}{399--408}.
\newblock


\bibitem[\protect\citeauthoryear{Rossi and Utkus}{Rossi and Utkus}{2020a}]%
        {rossi2020needs}
\bibfield{author}{\bibinfo{person}{Alberto~G Rossi} {and}
  \bibinfo{person}{Stephen~P Utkus}.} \bibinfo{year}{2020}\natexlab{a}.
\newblock \showarticletitle{The needs and wants in financial advice: Human
  versus robo-advising}.
\newblock \bibinfo{journal}{\emph{Available at SSRN 3759041}}
  (\bibinfo{year}{2020}).
\newblock


\bibitem[\protect\citeauthoryear{Rossi and Utkus}{Rossi and Utkus}{2020b}]%
        {rossi2020benefits}
\bibfield{author}{\bibinfo{person}{Alberto~G Rossi} {and}
  \bibinfo{person}{Stephen~P Utkus}.} \bibinfo{year}{2020}\natexlab{b}.
\newblock \showarticletitle{Who benefits from robo-advising? Evidence from
  machine learning}.
\newblock \bibinfo{journal}{\emph{Evidence from Machine Learning (March 10,
  2020)}} (\bibinfo{year}{2020}).
\newblock


\bibitem[\protect\citeauthoryear{Schumaker, Zhang, Huang, and Chen}{Schumaker
  et~al\mbox{.}}{2012}]%
        {schumaker2012evaluating}
\bibfield{author}{\bibinfo{person}{Robert~P Schumaker}, \bibinfo{person}{Yulei
  Zhang}, \bibinfo{person}{Chun-Neng Huang}, {and} \bibinfo{person}{Hsinchun
  Chen}.} \bibinfo{year}{2012}\natexlab{}.
\newblock \showarticletitle{Evaluating sentiment in financial news articles}.
\newblock \bibinfo{journal}{\emph{Decision Support Systems}}
  \bibinfo{volume}{53}, \bibinfo{number}{3} (\bibinfo{year}{2012}),
  \bibinfo{pages}{458--464}.
\newblock


\bibitem[\protect\citeauthoryear{Silva, Tabak, and Ferreira}{Silva
  et~al\mbox{.}}{2019}]%
        {silva2019modeling}
\bibfield{author}{\bibinfo{person}{Thiago~Christiano Silva},
  \bibinfo{person}{Benjamin~Miranda Tabak}, {and}
  \bibinfo{person}{Idamar~Magalh{\~a}es Ferreira}.}
  \bibinfo{year}{2019}\natexlab{}.
\newblock \showarticletitle{Modeling Investor Behavior Using Machine Learning:
  Mean-Reversion and Momentum Trading Strategies}.
\newblock \bibinfo{journal}{\emph{Complexity}}  \bibinfo{volume}{2019}
  (\bibinfo{year}{2019}).
\newblock


\bibitem[\protect\citeauthoryear{Sohangir, Wang, Pomeranets, and
  Khoshgoftaar}{Sohangir et~al\mbox{.}}{2018}]%
        {sohangir2018big}
\bibfield{author}{\bibinfo{person}{Sahar Sohangir}, \bibinfo{person}{Dingding
  Wang}, \bibinfo{person}{Anna Pomeranets}, {and} \bibinfo{person}{Taghi~M
  Khoshgoftaar}.} \bibinfo{year}{2018}\natexlab{}.
\newblock \showarticletitle{Big Data: Deep Learning for financial sentiment
  analysis}.
\newblock \bibinfo{journal}{\emph{Journal of Big Data}} \bibinfo{volume}{5},
  \bibinfo{number}{1} (\bibinfo{year}{2018}), \bibinfo{pages}{1--25}.
\newblock


\bibitem[\protect\citeauthoryear{Tauni, Yousaf, and Ahsan}{Tauni
  et~al\mbox{.}}{2020}]%
        {tauni2020investor}
\bibfield{author}{\bibinfo{person}{Muhammad~Zubair Tauni},
  \bibinfo{person}{Salman Yousaf}, {and} \bibinfo{person}{Tanveer Ahsan}.}
  \bibinfo{year}{2020}\natexlab{}.
\newblock \showarticletitle{Investor-advisor Big Five personality similarity
  and stock trading performance}.
\newblock \bibinfo{journal}{\emph{Journal of Business Research}}
  \bibinfo{volume}{109} (\bibinfo{year}{2020}), \bibinfo{pages}{49--63}.
\newblock


\bibitem[\protect\citeauthoryear{Thompson, Feng, Reesor, and Grace}{Thompson
  et~al\mbox{.}}{2021}]%
        {thompson2021know}
\bibfield{author}{\bibinfo{person}{John~RJ Thompson}, \bibinfo{person}{Longlong
  Feng}, \bibinfo{person}{R~Mark Reesor}, {and} \bibinfo{person}{Chuck Grace}.}
  \bibinfo{year}{2021}\natexlab{}.
\newblock \showarticletitle{Know Your Clients' behaviours: a cluster analysis
  of financial transactions}.
\newblock \bibinfo{journal}{\emph{Journal of Risk and Financial Management}}
  \bibinfo{volume}{14}, \bibinfo{number}{2} (\bibinfo{year}{2021}),
  \bibinfo{pages}{50}.
\newblock


\bibitem[\protect\citeauthoryear{Vanguard}{Vanguard}{2020}]%
        {vanguard2020invest}
\bibfield{author}{\bibinfo{person}{Vanguard}.} \bibinfo{year}{2020}\natexlab{}.
\newblock \showarticletitle{How America Invests: Retail research report}.
\newblock \bibinfo{journal}{\emph{Vanguard research}} (\bibinfo{year}{2020}).
\newblock
\newblock
\shownote{Available at retail.vanguard.com.}


\bibitem[\protect\citeauthoryear{Vanguard}{Vanguard}{2021}]%
        {vanguard2021save}
\bibfield{author}{\bibinfo{person}{Vanguard}.} \bibinfo{year}{2021}\natexlab{}.
\newblock \showarticletitle{How America Saves: A report on Vanguard 2020
  defined contribution plan data}.
\newblock \bibinfo{journal}{\emph{Vanguard research}} (\bibinfo{year}{2021}).
\newblock
\newblock
\shownote{Available at institutional.vanguard.com.}


\bibitem[\protect\citeauthoryear{Zhang and Wang}{Zhang and Wang}{2015}]%
        {zhang2015limited}
\bibfield{author}{\bibinfo{person}{Bing Zhang} {and} \bibinfo{person}{Yudong
  Wang}.} \bibinfo{year}{2015}\natexlab{}.
\newblock \showarticletitle{Limited attention of individual investors and stock
  performance: Evidence from the ChiNext market}.
\newblock \bibinfo{journal}{\emph{Economic Modelling}}  \bibinfo{volume}{50}
  (\bibinfo{year}{2015}), \bibinfo{pages}{94--104}.
\newblock


\end{thebibliography}
